\pdfoutput=1
\documentclass{article}

\usepackage{arxiv}

\usepackage[utf8]{inputenc} 
\usepackage[T1]{fontenc}    
\usepackage{hyperref}       
\usepackage{url}            
\usepackage{booktabs}       
\usepackage{amsfonts}       
\usepackage{nicefrac}       
\usepackage{microtype}      
\usepackage{lipsum}
\usepackage{moreverb,url}
\newcommand{\ignore}[1]{}
\usepackage{tikz}
\usepackage{subfigure}
\usepackage{textcomp}
\usepackage{amsmath,amsfonts,amssymb}
\usepackage[T1]{fontenc}
\usepackage{cite}
\usepackage[export]{adjustbox}

\title{Sensitivity-Enhanced Fourier Transform Mid-Infrared Spectroscopy Using a Supercontinuum Laser Source}

\author{Ivan Zorin\thanks{\textbf{Corresponding author:} \href{mailto:ivan.zorin@recendt.at}{ivan.zorin@recendt.at}} \\
	Research Center for Non-Destructive Testing\\
	Science Park 2, Altenberger Str.69\\
	4040 Linz, Austria \\
	\texttt{ivan.zorin@recendt.at} \\
	\And
	Jakob Kilgus \\
	Research Center for Non-Destructive Testing\\
	Science Park 2, Altenberger Str.69\\
	4040 Linz, Austria \\
	\And
	Kristina Duswald \\
	Research Center for Non-Destructive Testing\\
	Science Park 2, Altenberger Str.69\\
	4040 Linz, Austria \\
	\And
	Bernhard Lendl \\
	Institute for Chemical Technologies and Analytics\\
	TU Wien, Getreidemarkt 9\\
	1060 Vienna, Austria \\
	\And
	Bettina Heise \\
	Research Center for Non-Destructive Testing\\
	Science Park 2, Altenberger Str.69\\
	4040 Linz, Austria \\
	\And
	Markus Brandstetter \\
	Research Center for Non-Destructive Testing\\
	Science Park 2, Altenberger Str.69\\
	4040 Linz, Austria \\
	\texttt{markus.brandstetter@recendt.at} \\
}

\begin{document}
\maketitle

\vspace{-5pt}
\begin{abstract}
	\vspace{-5pt}

Fourier transform infrared (FTIR) spectrometers have been the dominant technology in the field of mid-infrared (MIR)spectroscopy for decades. Supercontinuum laser sources operating in the MIR spectral region now offer the potential to enrich the field of FTIR spectroscopy due to their distinctive properties, such as high-brightness, broadband spectral coverage and enhanced stability. In our contribution, we introduce this advanced light source as a replacement for conventional thermal emitters. Furthermore, an approach to efficient coupling of pulsed MIR supercontinuum sources to FTIR spectrometers is proposed and considered in detail. The experimental part is devoted to pulse-to-pulse energy fluctuations of the applied supercontinuum laser, performance of the system, as well as the noise and long-term stability.
Comparative measurements performed with a conventional FTIR instrument equipped with a thermal emitter illustrate that similar noise levels can be achieved with the supercontinuum-based system. The analytical performance of the supercontinuum-based FTIR spectrometer was tested for a concentration series of aqueous formaldehyde solutions in
a liquid flow cell (500~\textmu mm path length) and compared with the conventional FTIR (130~\textmu m path length). The results show a four-times-enhanced detection limit due to the extended path length enabled by the high brightness of the laser.
In conclusion, FTIR spectrometers equipped with novel broadband MIR supercontinuum lasers could outperform traditional systems providing superior performance, e.g., interaction path lengths formerly unattainable, while maintaining low noise levels known from highly stable thermal emitters.

\end{abstract}


\section{Introduction}
Fourier-transform infrared (FTIR) \ignore{spectrometers have}spectroscopy has been a well established and widely used tool for chemical characterization in various application scenarios for decades. FTIR spectrometers are still the gold standard in the mid-infrared (MIR) spectral range exhibiting reasonable acquisition times, high sensitivity and spectral resolution\cite{gunzler2002ir}. 
In their typical configuration, these spectrometers employ thermal light sources emitting black-body radiation perfectly fitted for many applications. Nevertheless, thermal emitters impose several limitations caused by their inherent properties, such as spatial incoherence, low power and omnidirectionality.
Recently, a highly interesting new type of broadband source has been emerging into the MIR spectral region, namely the supercontinuum laser source. 
MIR supercontinuum lasers are nowadays operating in the same or broader wavelength range as thermal sources\cite{dudley_taylor_2010,Dai_2018}. In contrast to \ignore{thermal sources}the latter, they exhibit drastically higher brightness, spatial coherence, and stability~\cite{PETERSEN2018182,Moselund:13} making them a promising tool for spectroscopy. Furthermore, supercontinuum sources are an attractive alternative to the already established MIR quantum cascade lasers (QCL)\cite{C7CS00403F,doi:10.1021/acs.analchem.8b01632}.

The noise characteristics of supercontinuum generation, which had been a deterrent in their early days, were significantly improved since then\cite{Newbury:03,1314060,Moselund:13}. 
Successful demonstrations of MIR supercontinuum source-based setups in a wide variety of noise sensitive applications \cite{KilgusApS:18,KilgusOE:18,Borondics:18,amiot,Gasser:18,Zorin:18} prove their practical suitability. Furthermore, it is expected that the trend of noise reduction will continue e.g. by implementing the supercontinuum generation in an all-normal-dispersion regime, which is insensitive to the input pump noise delivering the enhanced shot-to-shot spectral coherence\cite{Klimczak:16,Klimczak:14,Dupont_2014,Jiao:19,Genier:19}.
Meanwhile, with respect to the achieved average power, intensities up to 21.8~W were recently achieved in the wavelength range of 1.9~\textmu m~-~3.8~\textmu m\cite{Liu:14}.
The broad spectral range covered by these sources is continuously extending~\cite{Wang:17,Kubat:14,Singh:15,Cheng:16,16um,OB133} even beyond the fingerprint region revealing attractive potentials for spectroscopic measurements. This ultra-broadband coverage, including both near-infrared (NIR) and also the MIR range, is unique and only offered by supercontinuum lasers, while QCLs inherently offer a narrow-band wavelength tunability.

Due to the strict requirements on precise control of the mirror position, FTIR spectrometers are designed to operate with continuous-wave (CW) sources as the modulated signal is sampled at a defined frequency. This scheme introduces the limitations for operating FTIR spectrometers with pulsed sources without taking care of the sampling and pulse repetition rates. For this reason, the latter is usually set much higher to go into quasi-CW mode. This approach has been evaluated by e.g. the combination of an FTIR and a NIR supercontinuum source radiating at repetition rates of 80~MHz and 100~MHz already resulting in an improved signal-to-noise ratio (SNR) and detection limits as well as extended interaction path lengths\cite{Michaels:09,Goncharov,Mandon:08}.
In the much more attractive MIR spectral range, currently available supercontinuum sources typically operate at repetition rates from tens of kHz and up to several MHz with a pulse duration in the sub-nanosecond range. Due to the asynchronization of the emission and sampling performed by FTIR spectrometers, additional noise and spectral distortions are introduced resulting in insufficient stability of the signal for FTIR measurements\cite{Othman,Moselund16}. CW supercontinuum may be a key to solve the problem, however, the technical complexity and disadvantages (fibers with several tens of kilometers length, high power requirements, poor broadening)~\cite{dudley_taylor_2010,Nicholson2003} make them challenging especially in the MIR spectral range.  
From an analytical point of view, CW operation of such bright sources would introduce substantial thermal load, which may cause damage of the sample under investigation. 

In this contribution, we prove the principle suitability of pulsed MIR supercontinuum lasers as a light source in an FTIR instrument. We define and evaluate practical problems for coupling with an FTIR spectrometer that is highly sensitive to intensity variations and report a successful method to overcome the sampling problem. In the experimental part, we examine pulse-to-pulse intensity fluctuations over the supercontinuum emission spectrum, the noise of the experimental setup and its stability. Finally, the analytical performance of the proposed solution is investigated and compared to conventional state-of-the-art instrumentation. 

\section{Materials and methods}

\subsection{Mid-infrared supercontinuum laser source}
In our study, a novel and commercially available supercontinuum source (NKT Photonics, SuperK MIR) with a spectral coverage from 1.1~\textmu m to 4.4~\textmu m was applied. The ultra-broadband spectrum is generated due to a sequence of non-linear processes initiated in a ZBLAN (ZrF\textsubscript{4}-BaF\textsubscript{2}-LaF\textsubscript{3}-AlF\textsubscript{3}-NaF) fiber\cite{dudley_taylor_2010}, pumped in the spectral region of around 2~\textmu m by a pre-broadened and amplified seed laser (1.55~\textmu m). 

The repetition rate of the source is adjustable in the range of 2~MHz~to~3~MHz (2.5~MHz default). 
The supercontinuum pulse width is shorter than 1~ns yielding in a duty cycle of less than 0.5\% causing the spectrum distortion and irreproducibility of FTIR measurements when directly applied without additional measures\cite{Othman}. 
The output beam of the supercontinuum source is a single Gaussian mode with a beam quality parameter $\mathrm{M^2\leq1.1}$ and a diameter of around 4~mm with a beam divergence of $2$~mrad. The total average power of the source was measured as 475~mW, whereof 200~mW are radiated in MIR spectral range.

Characterization and verification of the pulse-to-pulse energy fluctuations were performed using a monochromator (Gilden Photonics, GM500) and a high-speed Mercury Cadmium Telluride (MCT) detector (PVM series, Vigo, rise time $\leq0.7$~ns and 230~MHz bandwidth).

\subsection{Experimental setup}
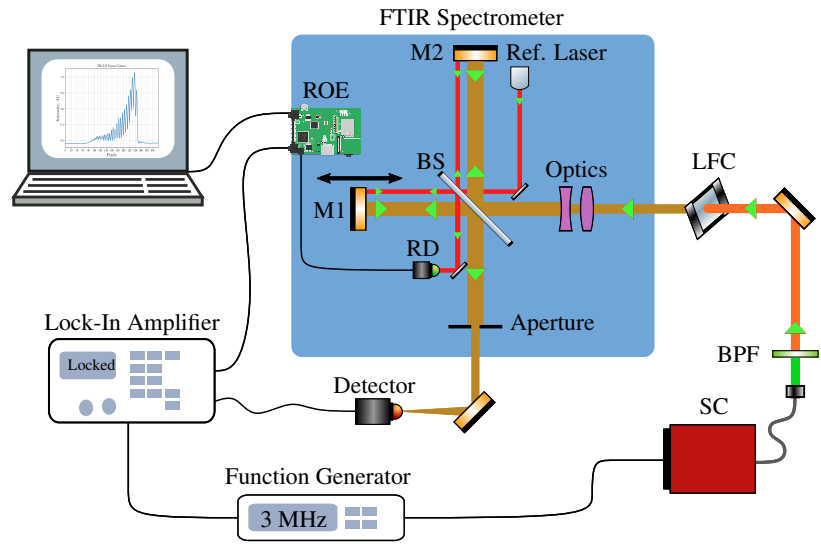
\begin{figure}[ht]
\centering
\begin{tikzpicture}
  \node[anchor=south west,inner sep=0] (image) at (0,0,0) {\includegraphics[width=0.65\linewidth]{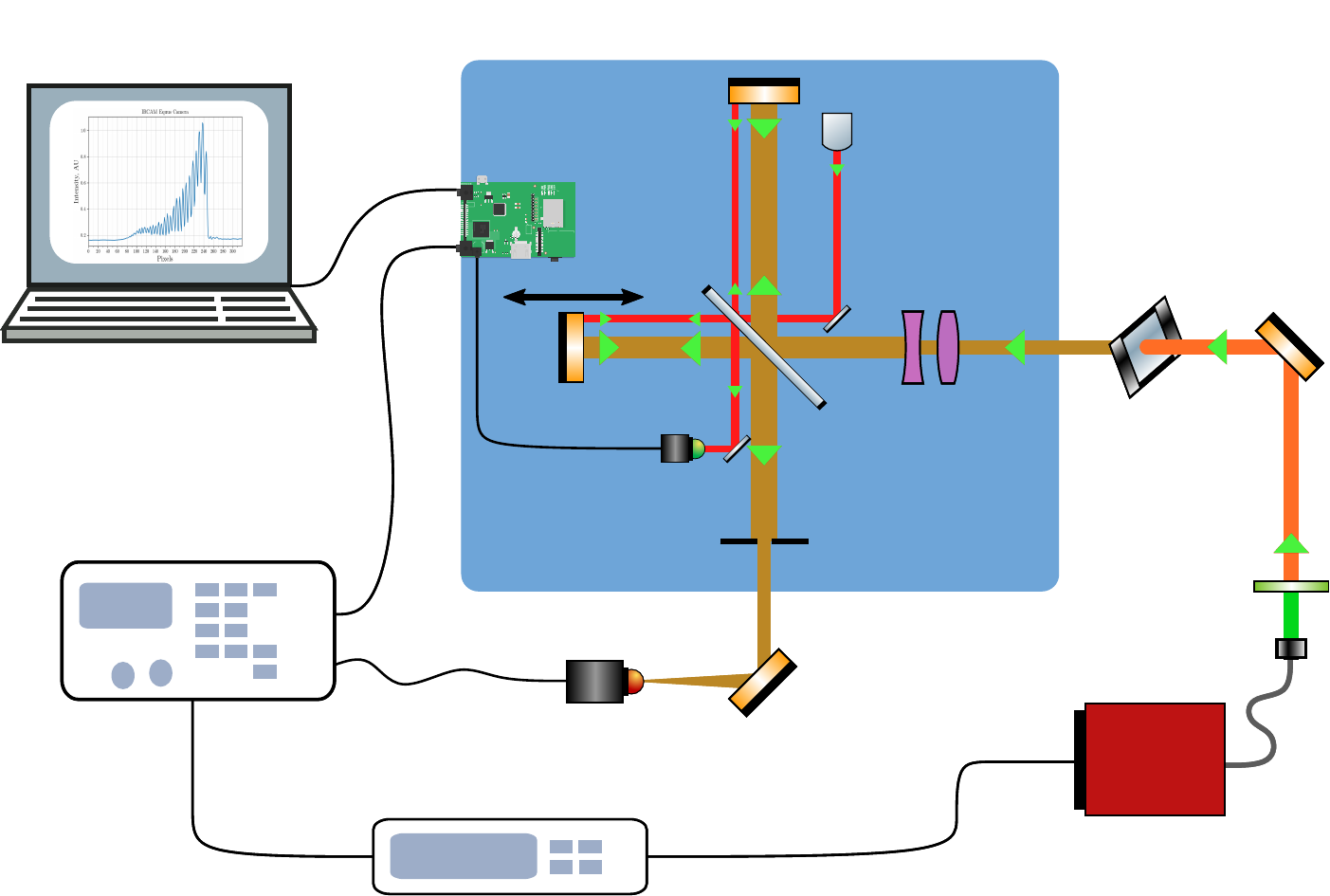}};
  \begin{scope}[x={(image.south east)},y={(image.north west)}]

    \draw (0.38,0.12) node[]{\color{black}\footnotesize Function Generator};
    \draw (0.15,0.4) node[]{\color{black}\footnotesize Lock-In Amplifier};
    \draw (0.87,0.25) node[]{\color{black}\footnotesize SC};
    \draw (0.87,0.7) node[]{\color{black}\footnotesize LFC};
    \draw (0.7,0.68) node[]{\color{black}\footnotesize Optics};
    \draw (0.395,0.61) node[]{\color{black}\footnotesize M1};
    \draw (0.515,0.9) node[]{\color{black}\footnotesize M2};
    \draw (0.675,0.9) node[]{\color{black}\footnotesize Ref. Laser};
    \draw (0.52,0.7) node[]{\color{black}\footnotesize BS};
    \draw (0.67,0.4) node[]{\color{black}\footnotesize Aperture};
    \draw (0.45,0.287) node[]{\color{black}\footnotesize Detector};
    \draw (0.51,0.54) node[]{\color{black}\footnotesize RD};
    \draw (0.35,0.042) node[]{\color{black}\footnotesize 3 MHz};
    \draw (0.1,0.325) node[]{\color{black}\tiny Locked};
    \draw (0.9,0.345) node[]{\color{black}\footnotesize BPF};
    \draw (0.39,0.83) node[]{\color{black}\footnotesize ROE};

    \draw (0.57,0.96) node[]{\color{black}\footnotesize FTIR Spectrometer};

  \end{scope}
\end{tikzpicture}
\caption{Scheme of the experimental setup: FTIR spectrometer~(Bruker Vertex~70, simplified scheme), SC~-~supercontinuum laser source, BPF~-~band-pass spectral filter, LFC~-~liquid flow cell, ROE~-~spectrometer built-in read-out electronics and mirror control, M1 and M2~-~interferometer mirrors and BS~-~beam splitter; RD~-~detector used to produce reference interferogram.}
\label{fig:scheme}
\end{figure}

The experimental setup consisted of a commercial FTIR instrument (Bruker Optics, Vertex~70) and the pulsed supercontinuum laser as depicted in Fig.~\ref{fig:scheme}.

In order to effectively exploit the dynamic range of the detector and avoid oversaturation by NIR spectral components (e.g. strong and high power seed laser line at 1.55~\textmu m), a suitable bandpass filter (BPF, Thorlabs FB3500-500,  3100~cm\textsuperscript{-1} - 2650~cm\textsuperscript{-1}) was employed thereby determining the investigated spectral range; the average power of the transmitted band (i.e. incident on the sample) was measured equal to~17~mW (power meter, Coherent, LM-10 HTD), which did not lead to observable heating of the sample. 
It should be noted that the measurements over the entire MIR part of the emission spectrum are restricted by the dynamic range of the detector, however, they are realizable applying edgepass (1.65~\textmu m cut-on wavelength) and neutral density filters or introducing smaller diameters of the aperture.

The collimated radiation was transmitted through the sample inserted in a liquid flow cell (PIKE Technologies, 4~mm CaF\textsubscript{2} windows) before it entered the FTIR spectrometer via the external optical input.

The FTIR spectrometer with a typical configuration based on a Michelson configuration (BS - beam splitter, M1 and M2 - movable and fixed mirrors, respectively) is schematically shown in~Fig.~\ref{fig:scheme}. The interferogram was recorded with a Mercury Cadmium Telluride detector (variable gap Vigo PCI-4TE-12, detectivity within the spectral range D$\geq~2.0\times~10^{9}$~$\mathrm{cm\cdot \sqrt{Hz}\cdot W^{-1}}$).

\begin{figure}[ht]
\centering
\begin{tikzpicture}
  \node[anchor=south west,inner sep=0] (image) at (0,0,0) {\includegraphics[width=0.65\linewidth]{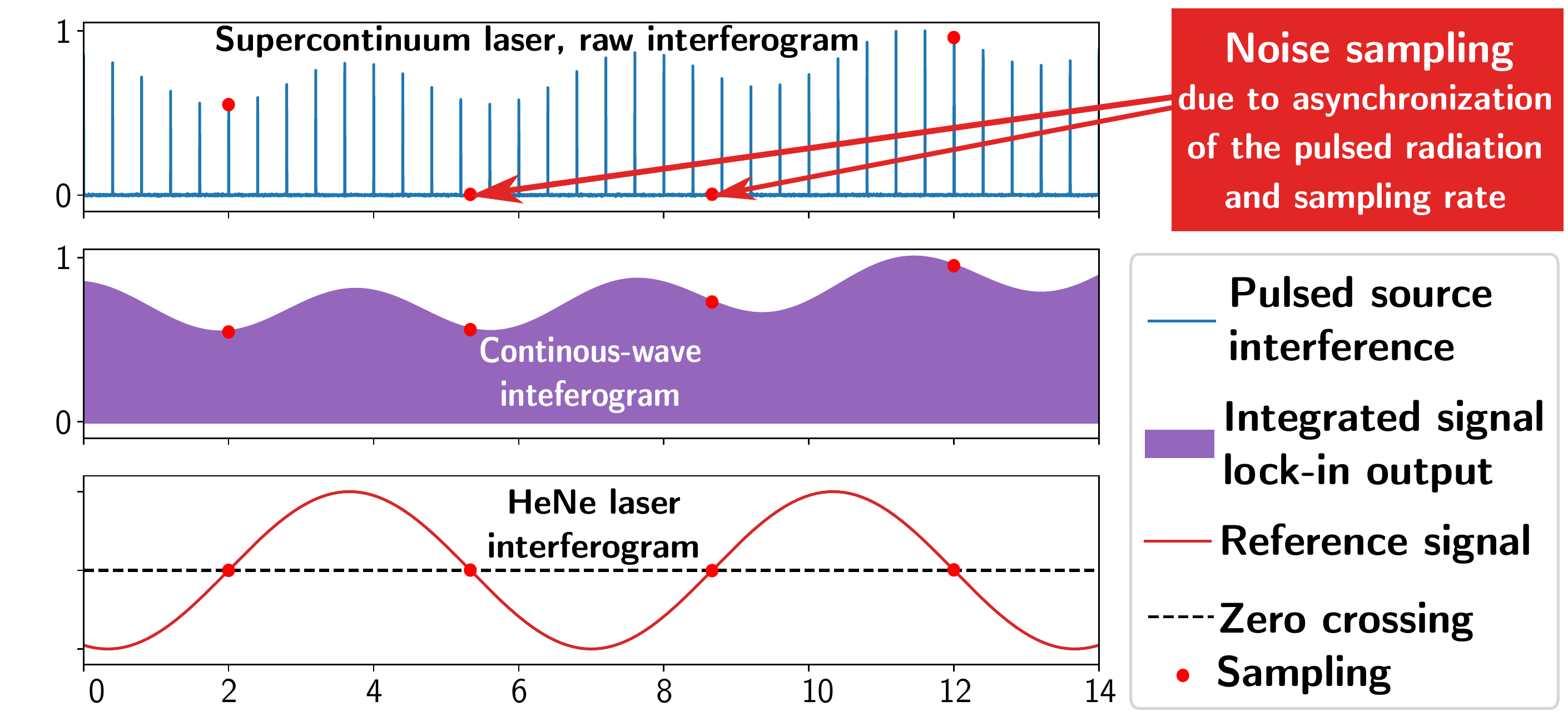}};
  \begin{scope}[x={(image.south east)},y={(image.north west)}]

    \draw (0.38,-0.05) node[]{\color{black} Time (\textmu s)};
    \draw (-0.02,0.5) node[rotate=90]{\color{black} Intensity (a.u.)};
    \draw (0.025,0.84) node[]{\color{black}\footnotesize a)};
    \draw (0.025,0.52) node[]{\color{black}\footnotesize b)};
    \draw (0.025,0.22) node[]{\color{black}\footnotesize c)};

    \draw[dashed,line width=0.8pt, orange] (0.146,0.22) -- (0.146,0.85);
    \draw[dashed,line width=0.8pt, orange] (0.3,0.22) -- (0.3,0.725);
    \draw[dashed,line width=0.8pt, orange] (0.455,0.22) -- (0.455,0.725);
    \draw[dashed,line width=0.8pt, orange] (0.608,0.22) -- (0.608,0.95);

  \end{scope}
\end{tikzpicture}
\caption{Problem of the signal sampling for low-duty cycle sources applied in the FTIR spectrometer and approach to overcome the asynchronization using an external integrator: (a)~Interferogram for the pulsed light source, (b) Integrated CW signal of the detector, i.e. lock-in amplifier (LIA) output, (c)~Reference interferogram produced by the monochromatic source, defines the sampling frequency (red).}
\label{fig:problem}
\end{figure}
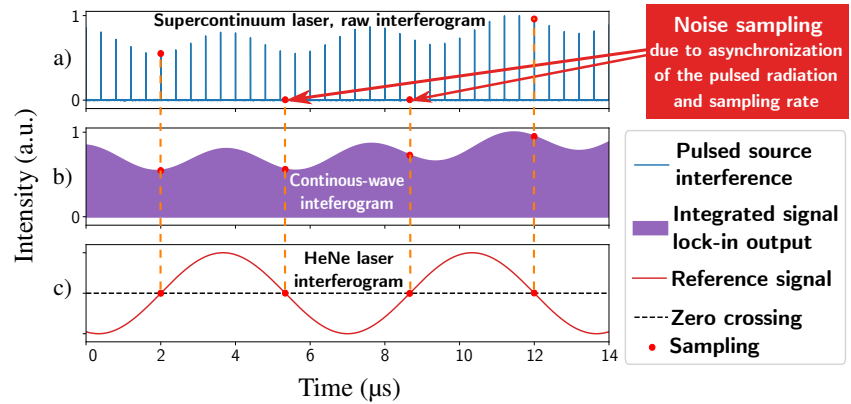

The built-in spectrometer control and read-out unit (ROE) utilized for the signal acquisition was used to control mirror scan velocity (i.e sampling frequency) and size of the output aperture. In order to achieve the equidistant time sampling, the reference monochromatic laser is used to produce a quasi-sine interferogram~\cite{griffiths2007fourier} (e.g. see~Fig.~\ref{fig:problem}(c)) to be recorded by the reference detector (RD).

\begin{figure}[ht]
\centering
\begin{tikzpicture}
  \node[anchor=south west,inner sep=0] (image) at (0,0,0) {\includegraphics[width=0.7\linewidth]{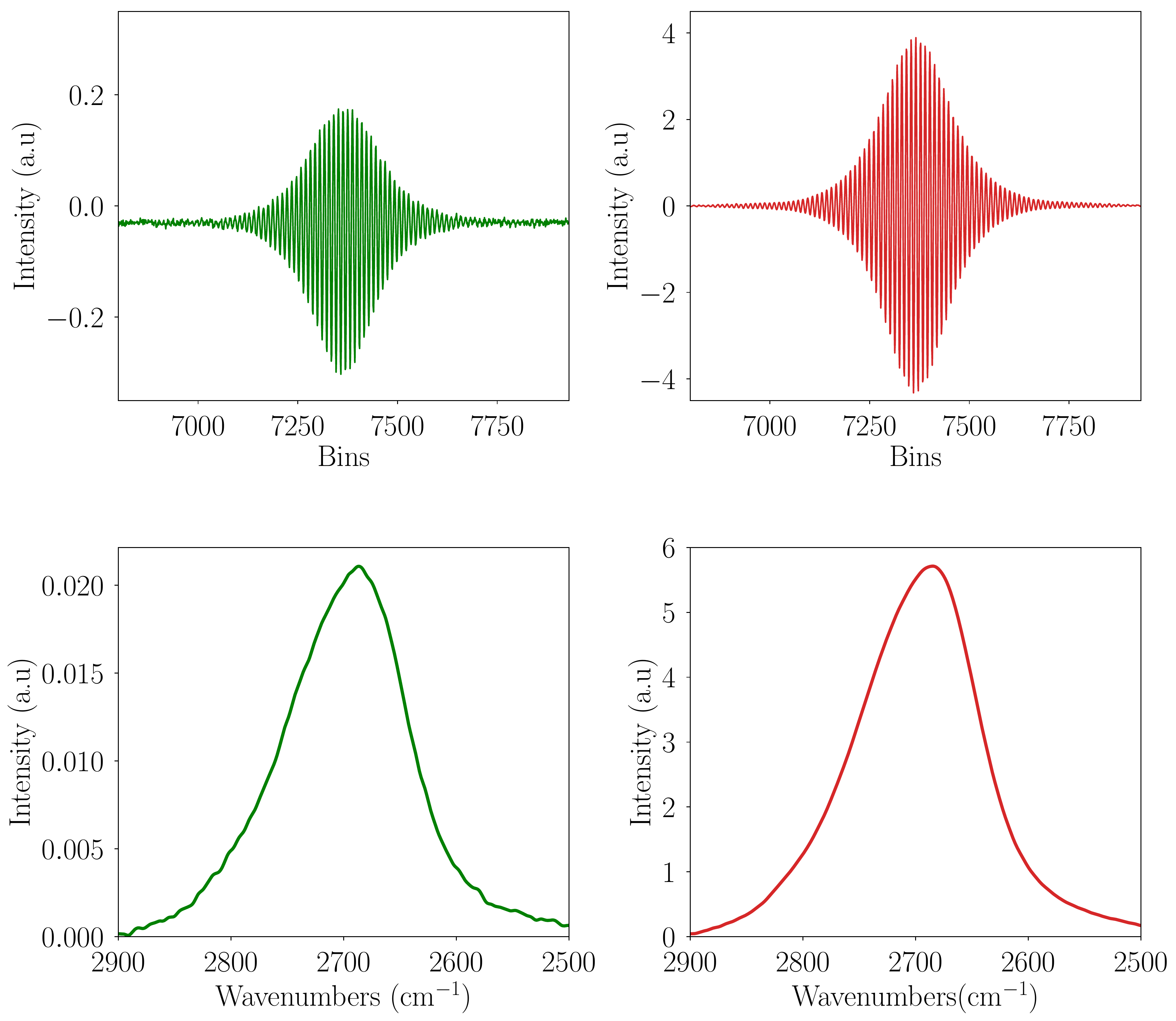}};
  \begin{scope}[x={(image.south east)},y={(image.north west)}]

    \draw (0.15,0.95) node[]{\color{black} a)};
    \draw (0.15,0.425) node[]{\color{black} b)};
    \draw (0.625,0.95) node[]{\color{black} c)};
    \draw (0.625,0.425) node[]{\color{black} d)};

    \draw (0.4,0.425) node[fill=orange!30,rectangle,rounded corners=2]{\color{black} \tiny \textbf {SNR=$\mathbf{3.74}$}};      
    \draw (0.89,0.425) node[fill=blue!30,rectangle,rounded corners=2]{\color{black} \tiny \textbf {SNR=$\mathbf{7320}$}};


  \end{scope}
\end{tikzpicture}
\caption{Raw interferograms and reconstructed emission spectra of the supercontinuum source (band-pass filter inserted) obtained using the traditional sampling approach without external integrator (a,b) and applying the lock-in amplifier (c,d). The intensity values are not normalized and correspond to the amplitude of the raw signals.}
\label{fig:sublockin}
\end{figure}
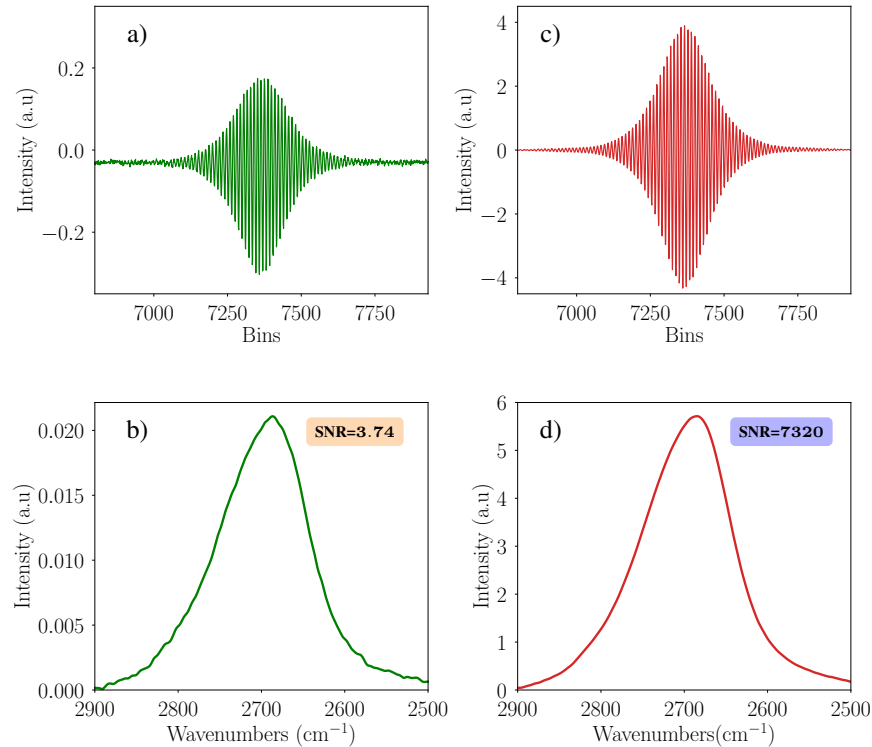

Any asynchronization of the signal acquisition and pulse generation would result in strong noise and low intensity of the recorded interferogram~(Fig.~\ref{fig:sublockin}), since the sampling would frequently be performed in the absence of any pulse, thereby being equivalent to the measurement of the detector noise. Figure~\ref{fig:problem} shows a zoomed-in part of a simulated interferogram and serves purely for illustration purposes defining the sampling problem and a possible solution by using an external integrator. The pulses presented in Fig.~\ref{fig:problem}(a) were chosen according to the parameters of the supercontinuum radiation exploited in our study. A reference signal (150~kHz, Fig.~\ref{fig:problem}(c)) that defines the sampling frequency was simulated in order to temporarily scale the graph and show the infeasibility of the traditional approach to sample the signal produced by any low-duty cycle source.

In the system proposed in~Fig.~\ref{fig:scheme}, the lock-in amplifier (LIA, Stanford Research Systems, SR844) was used as an integrator to suppress the pulsed nature of the signal. Therefore, the output of the MCT detector was transformed into a signal similar to that obtained in the traditional configuration with a CW source~(i.e. to one shown Fig.~\ref{fig:problem}(c)).
\ignore{In practice, the method is equivalent to modifying the detector time constant while maintaining the same high sensitivity.} According to the Nyquist-Kotelnikov-Shannon theorem, the measurement approach is feasible only for integration times of the LIA shorter than half of the period of sampling, otherwise, the interforogram is distorted or undetectable. Nevertheless, very short time constants that are achievable with modern LIAs (e.g. 1~\textmu s) result in \ignore{$\sqrt{t}$} a higher spectral noise\ignore{, where $t$ is a time constant,} due to the averaging of fewer pulses.

The LIA that was used in this work has a fixed and discrete integration time range (100~\textmu s, 300~\textmu s, 1~ms and longer). The sampling rate, i.e. the mirror velocity of the FTIR spectrometer, could be adjusted within the range from 1~kHz to 60~kHz. Therefore, the requirements defined by the sampling theorem were satisfied by selecting the proper parameters of the LIA and the FTIR spectrometer. With respect to the trade-off between the speed and sensitivity (100~\textmu s and 300~\textmu s time constants correspond to 5~kHz and 1.7~kHz sampling rates), the following measurements were executed at the longer integration time, i.e. 300~\textmu s, and at 1~kHz sampling frequency. The external function generator was utilized to trigger the supercontinuum source at the highest available frequency (3~MHz, according to the maximum repetition rate) and to synchronize the laser with the LIA. Consequently, around 900~pulses were integrated for each sampling point.

Figure~\ref{fig:sublockin} illustrates raw signals and spectra obtained by the supercontinuum-based FTIR spectrometer in two configurations: a direct coupling and a coupling using the LIA locked at the repetition rate of the source. The optical parameters and conditions were set the same for both measurements (aperture 8~mm, 500~\textmu m path length liquid flow cell, distilled water). SNR values were traditionally derived as the ratio between maximum signal levels and calculated standard deviations of the zero absorbance lines (also known as 100\% lines). It should be noted that the peak intensity when using the LIA-based integration scheme showed a 250~times higher magnitude, which is proportional to the pulse sampling probability, i.e. $1/\mathrm{D_c}$, where D\textsubscript{c} is the duty cycle.

\subsection{Samples}
Formaldehyde (CH\textsubscript{2}O or methanal) is a pollutant that is widely distributed in the environment. Due to its high water solubility~\cite{ALLOU20112991} and a wide variety of natural sources such as e.g. oxidation of organic matters\cite{glaze} and industrial effluents, formaldehyde and formaldehyde monohydrate (methanediol) can e.g. be frequently found in food, natural- and drinking water\cite{WHO_F}.

The evidence of mutagenicity and risks of carcinogenicity of formaldehyde are well investigated and verified\cite{Swenberg}. Histopathological short and long-term effects in the mucosa of rats are documented at the concentration threshold of 260~ppm\cite{TIL198977}, while a tolerable concentration of 2.6~ppm with an uncertainty factor of 100 is defined by the World Health Organization (WHO)\cite{WHO_F}.
During the metabolism following oral exposure, formaldehyde is oxidized to formic acid that also introduces specific toxic effects on human health\cite{toxic,Sadun423}.

The detection of low concentrations of formaldehyde dissolved in water is a practically interesting problem for cost-efficient spectroscopy since it is usually done by sophisticated and expensive chromatographic analysis\cite{YASRI2015487}.
In our study, a concentration series of formaldehyde aqueous solutions has been prepared for spectroscopic analysis by dilution from 37\% stock solution (Carl Roth GmbH).

\section{Results and discussions}

\begin{figure}[ht]
\centering
\subfigure[Emission spectra of the sources]{
\begin{tikzpicture}[remember picture]
  \node[anchor=south west,inner sep=0] (image) at (0,0,0) {\includegraphics[width=0.48\linewidth]{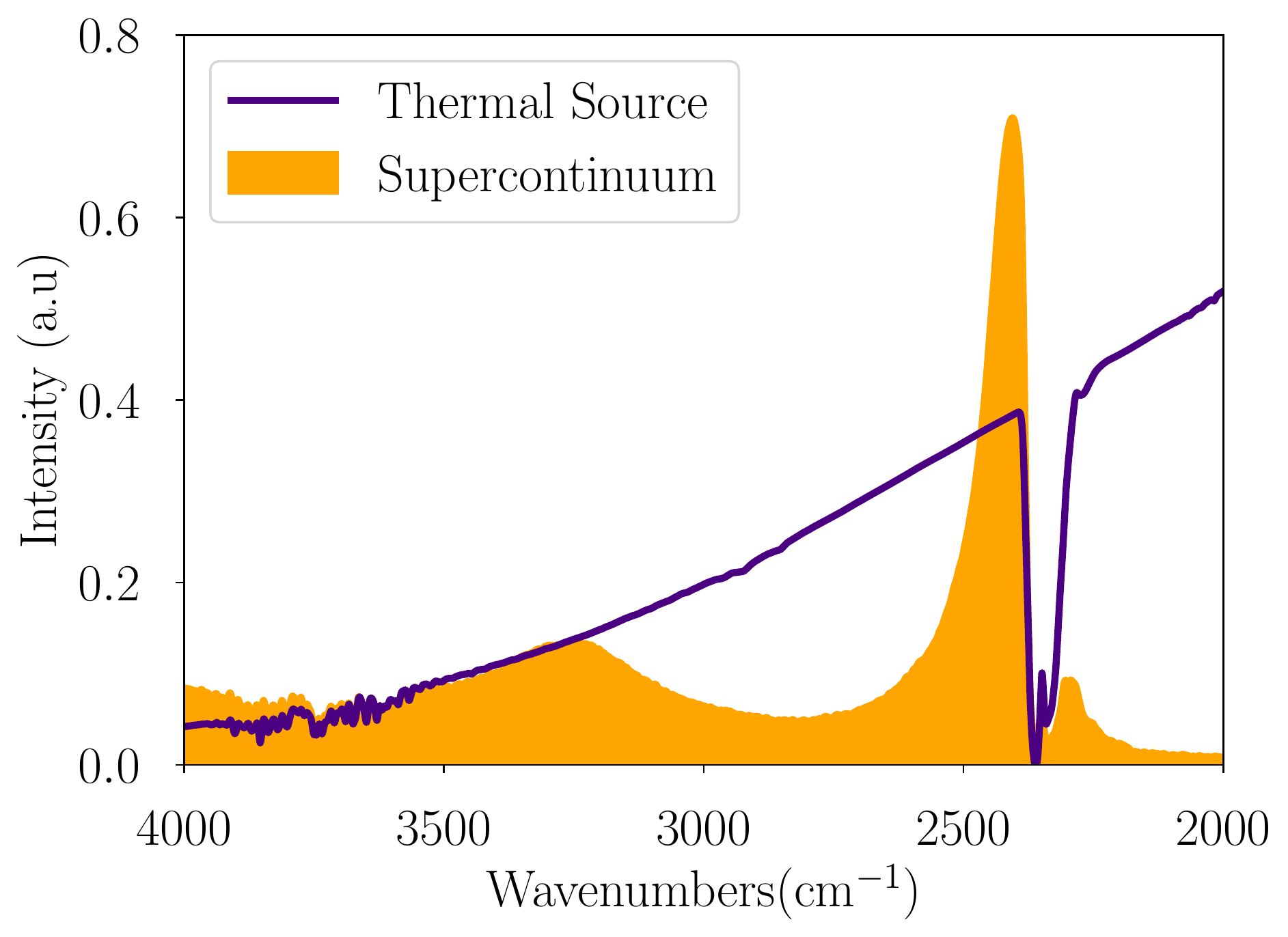}};
  \begin{scope}[x={(image.south east)},y={(image.north west)}]
  \end{scope}
\end{tikzpicture}
\label{fig:sour}}
 \subfigure[Absorption spectra of formaldehyde and water]{
 \begin{tikzpicture}[]
  \node[anchor=south west,inner sep=0] (image) at (0,0,0) {\includegraphics[width=0.48\linewidth]{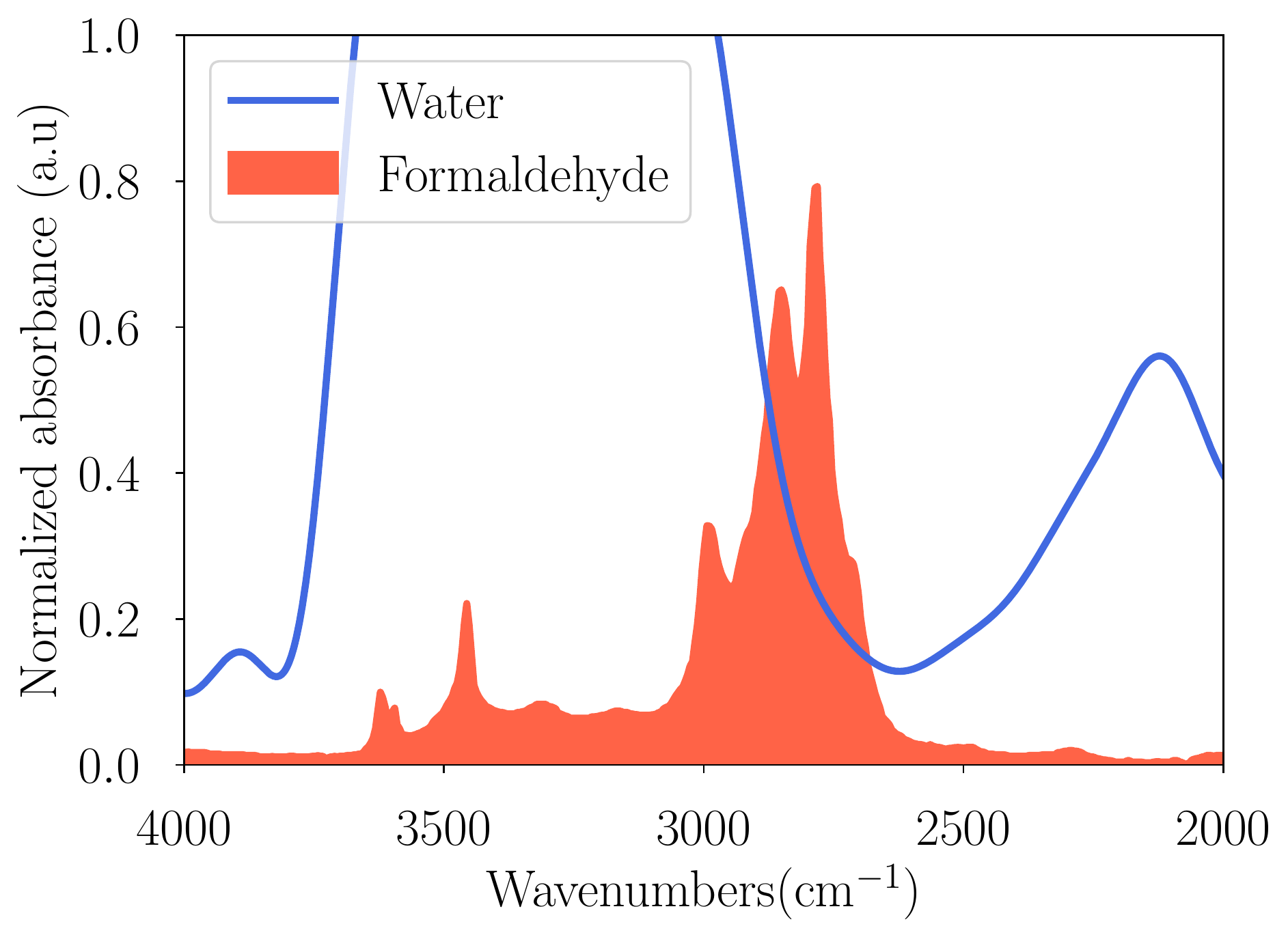}};
  \begin{scope}[x={(image.south east)},y={(image.north west)}]
  \end{scope}
\end{tikzpicture}
 \label{fig:absorb}}
    \begin{tikzpicture}[overlay,thick]
    \end{tikzpicture}
\caption{(a) Emission spectra of the supercontinuum laser source and the thermal source with respect to (b) absorption of water (recorded with the thermal emitter, 90~\textmu m liquid flow cell) and formaldehyde (spectrum taken from database\cite{nist_ir}). The supercontinuum spectrum was recorded using a neutral-density filter (optical density of 1), an aperture size was set to 0.25~mm; the emission spectrum of the thermal source was measured with an aperture size of 8~mm in the absence of any filters; both recorded emission spectra reveal strong absorption of CO\textsubscript{2} around 2350~cm\textsuperscript{-1}.}
\label{fig:spectra}
\end{figure}

The measured emission spectra of the thermal and supercontinuum source are depicted in~Fig.~\ref{fig:sour}, edgepass filter (1.65~\textmu m cut-on wavelength) and neutral density filter were used to avoid oversaturation of the detector. 

Intensity values are not normalized, since different optical parameters were used to avoid oversaturation of the MCT detector: the aperture was set to 8~mm diameter for the thermal emitter, while for the high brightness supercontinuum laser the aperture was set to the minimum available aperture of 0.25~mm with an additionally inserted neutral density filter (10\% transmission).

In order to determine the analytical performance (i.e. limit of detection) of the supercontinuum-based FTIR, an aqueous dilution series was measured. Formaldehyde, i.e more accurately its mixture with the hydrated form methanediol (CH\textsubscript{2}(OH)\textsubscript{2})\cite{MATUBAYASI200758} was chosen as the target analyte. Formaldehyde exhibits strong absorption due to the stretching vibration of C-H, in a spectral range where water absorption shows a relative minimum (see~Fig.~\ref{fig:absorb}, spectrum of formaldehyde taken from a database\cite{nist_ir}). Hence, the bandpass filter with a quasi-Gaussian profile was used to select the spectral range (3100~cm\textsuperscript{-1} - 2650~cm\textsuperscript{-1}); the default spectral resolution was set to 4~cm\textsuperscript{-1} for the following experimental verification and noise analysis.

\subsection{Noise and long-term stability}

\begin{figure}[ht]
\begin{center}
 \subfigure[100\% zero absorbance lines, 130~\textmu m liquid flow cell]{
 \begin{tikzpicture}[]
  \node[anchor=south west,inner sep=0] (image) at (0,0,0) {\includegraphics[width=0.48\linewidth]{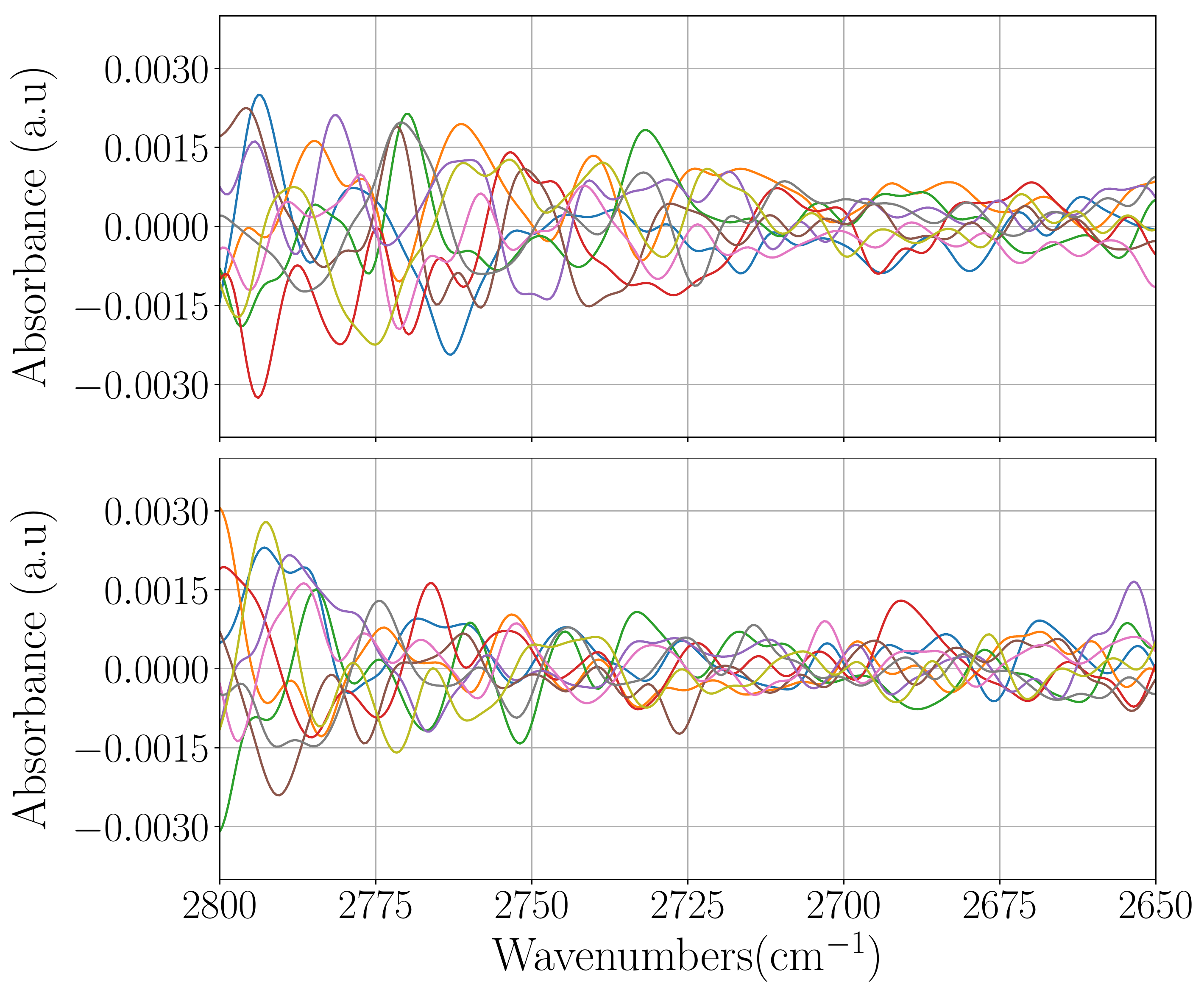}};
  \begin{scope}[x={(image.south east)},y={(image.north west)}]
  \end{scope}
\end{tikzpicture}
 \label{fig:huline}}
\subfigure[Allan-Werle variance]{
\begin{tikzpicture}[remember picture]
  \node[anchor=south west,inner sep=0] (image) at (0,0,0) {\includegraphics[width=0.48\linewidth, padding =0 24pt 0 0]{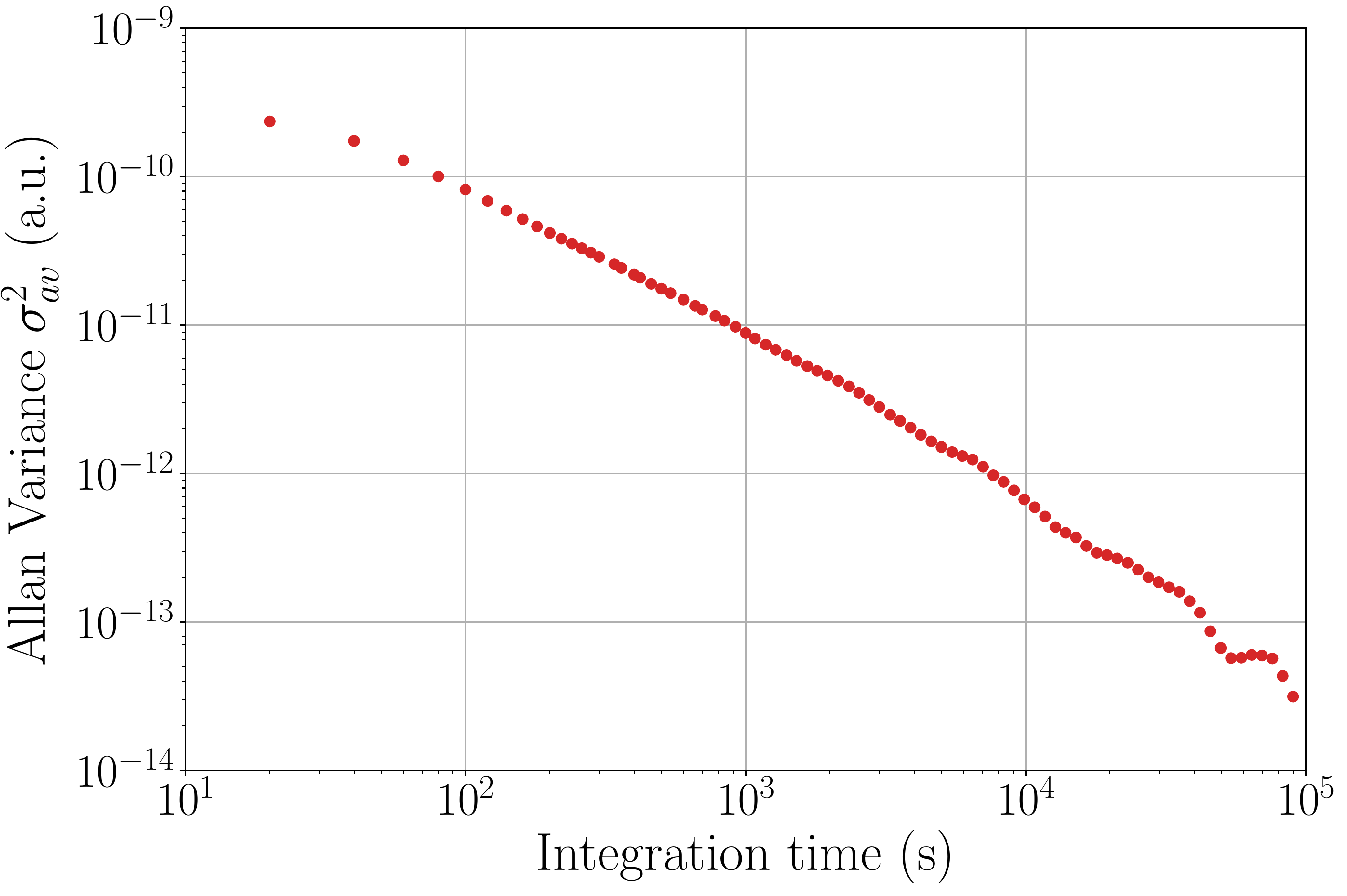}
};
  \begin{scope}[x={(image.south east)},y={(image.north west)}]
  \end{scope}
\end{tikzpicture}
\label{fig:allan}
}
    \begin{tikzpicture}[overlay,thick]
 	   \draw (-11.5,6-0.2) node[fill=orange!80,circle]{\color{white}\footnotesize \textbf{SC}};
 	   \draw (-11.5,3) node[fill=violet!60,circle]{\color{white}\footnotesize \textbf{TE}};
 	   \draw (-10,6-0.2)  node[]{\color{black} \footnotesize RMS=0.00078};
       \draw (-10,3)  node[]{\color{black} \footnotesize RMS=0.00067};
    \end{tikzpicture}
\caption{(a) Noise performance of the experimental setup using a 130 µm liquid flow cell filled with blank water: the supercontinuum source (marked as SC) in comparison to the thermal emitter (TE) expressed in the form of 100\% (zero absorbance) lines and (b) long-term stability of the LIA based setup illustrated by the Allan Werle variance.}
\label{fig:noise}
\end{center}
\end{figure}
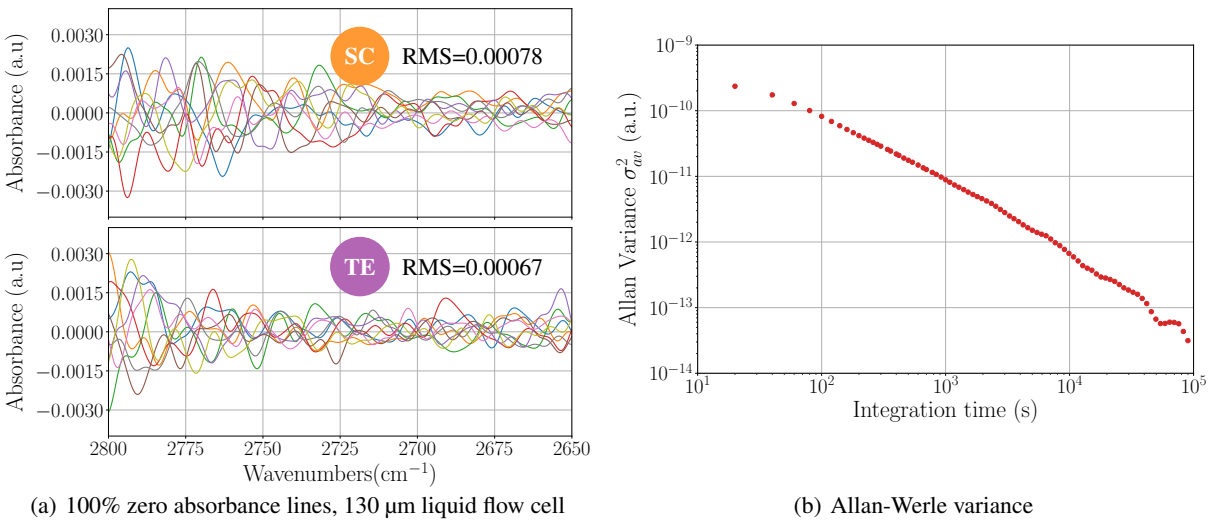

In order to characterize and compare the noise of the supercontinuum-based system (Fig.~\ref{fig:noise}) and conventional FTIR with the CW thermal emitter, zero absorbance lines were obtained within the spectral range of interest (3100~cm\textsuperscript{-1} - 2650~cm\textsuperscript{-1}).

 Since the spectroscopic analysis could not be performed directly for the same path length due to the different intensity levels of the thermal and supercontinuum source and the limited dynamic range of the detector\cite{Brandstetter2013}, a direct comparison of the same sample at the same optical path length was not feasible. Therefore, a neutral density filter (optical density of 1) was used to reduce the output power of the laser. Figure~\ref{fig:huline} depicts the recorded 100\% lines measured through a 130~\textmu m liquid flow cell\ignore{ (8~mm aperture)} filled with blank water. Four consecutive spectra were averaged for the calculation of each zero absorption line. Root-mean-square (RMS) values for both sources exhibit comparable noise levels (0.67$\times$10\textsuperscript{-3} and 0.78$\times$10\textsuperscript{-3} standard errors for thermal and supercontinuum correspondingly), while the raw signal level of the quasi-Gaussian peak transmitted through BPF shows a factor of 25 higher magnitude in the case of the supercontinuum laser\ignore{, i.e. 1.7 a.u. (arbitrary units) against 0.066 a.u}. The spectral region beyond 2800~cm\textsuperscript{-1} is not used in the calculations due to the total attenuation imposed by water and thus insufficient light transmitted in this range.

Additionally, long-term measurements using the same 130~\textmu m liquid flow cell (blank water) were carried out over around 28~hours (10.000 spectra) to specify the stability of the system, the prevalent types of noise sources and to estimate detection limits that could be achieved by averaging. The series of error values calculated for zero absorbance lines (within the same spectral range of 2800~cm\textsuperscript{-1} - 2650~cm\textsuperscript{-1}) were used to derive the Allan-Werle variance\cite{Werle1993} as depicted in~Fig.~\ref{fig:allan}.  
The dependence of the signal deviation on the integration time illustrated by the Allan-Werle variance demonstrates the stability of the supercontinuum-based FTIR in a long-term perspective. 

Analysing the plot, white noise can be observed as a dominant noise floor in the system, as indicated by the decreasing tendency with a constant slope. This noise is common and most likely inherited by the nonlinear processes (initiated by amplification of the pump shot noise)~\cite{Newbury:03,1314060}\ignore{\cite{Newbury:03,1314060,RevModPhys.78.1135}} that occurred during the spectral broadening. A weak long-term drift was observed only after 4$\times$10\textsuperscript{4} seconds and could be caused by temperature changes. 

The evolution of the variance reveals that the SNR of the system and the corresponding limit of detection could be enhanced by 4 orders of magnitude applying the signal averaging. However, such long averaging is not expedient so that the spectroscopic measurements in the following sections are performed with an averaging over 10 spectra only (around 70~seconds).

\subsection{Pulse-to-pulse energy fluctuations}

\begin{figure}[ht]
\centering
\begin{tikzpicture}
  \node[anchor=south west,inner sep=0] (image) at (0,0,0) {\includegraphics[width=0.49\linewidth]{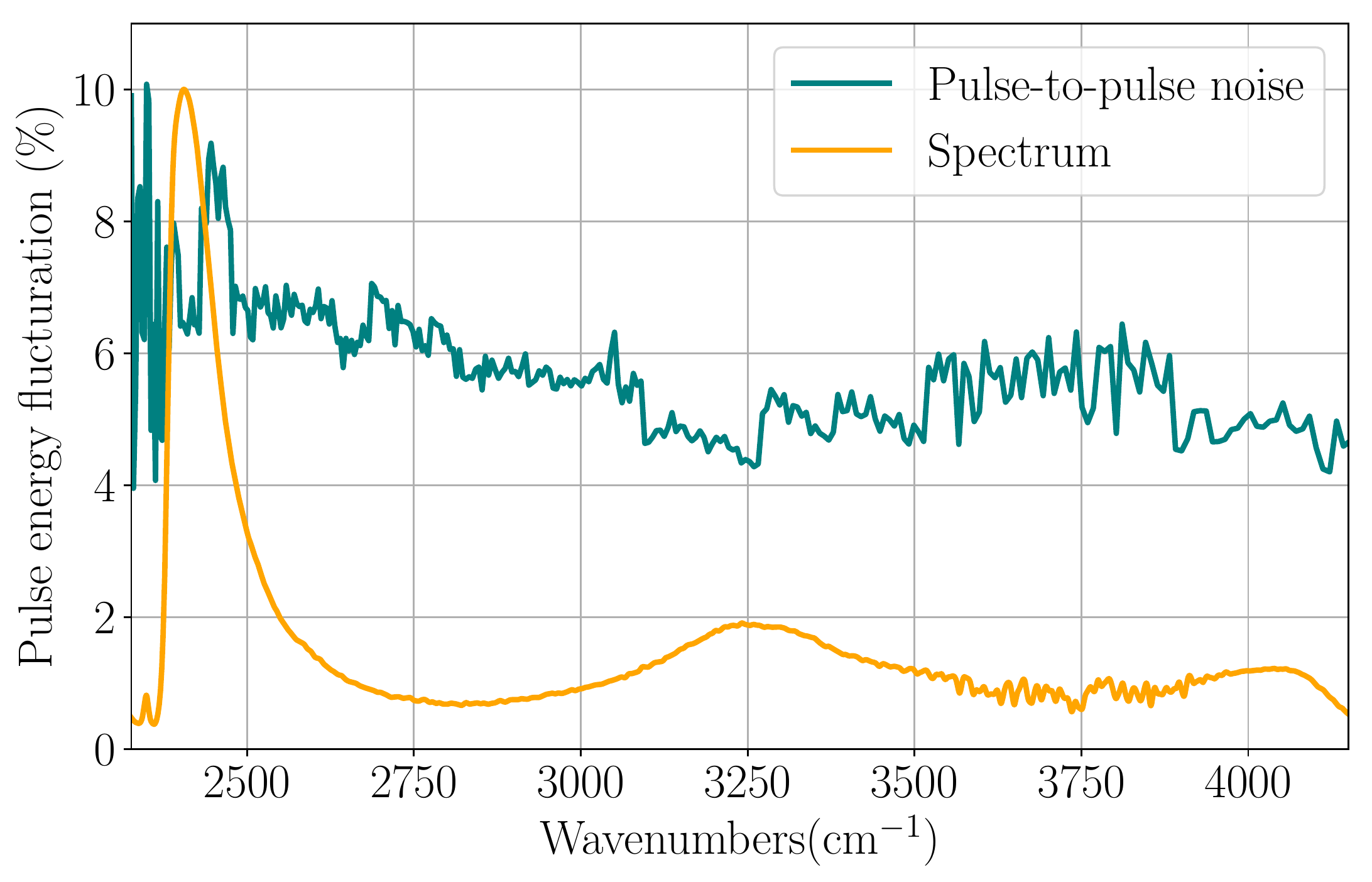}};
  \begin{scope}[x={(image.south east)},y={(image.north west)}]


  \end{scope}
\end{tikzpicture}
\caption{Normalized pulse-to-pulse energy fluctuations over the emission spectrum of the supercontinuum laser source (indicated for reference).}
\label{fig:pulse_fluct}
\end{figure}

To specify and complete the noise characterization of the supercontinuum source, the pulse-to-pulse behaviour over the emission spectrum has been investigated (Fig.~\ref{fig:pulse_fluct}). 

A Czerny-Turner monochromator \ignore{(Gilden Photonics GM500) }was used to access the pulse-to-pulse energy fluctuations within a narrow spectral band (3~nm resolution, equidistant in the wavelength space).
The high-speed MCT detector \ignore{(PVM series, Vigo, rise time $\leq0.7$~ns and 230~MHz bandwidth)}and oscilloscope (400 MHz, 5 GS/s) were used to record pulse waveforms. The measurements were performed by sweeping over the MIR part of the spectrum with 6~nm step size, while for each position 1000~pulses were analysed. The normalized pulse energy fluctuations, depicted in~Fig.~\ref{fig:pulse_fluct}, were calculated as a standard deviation of the pulse areas (time-integrals, represent a pulse energy) normalized by their mean.

An average pulse-to-pulse energy fluctuation of around 6.4\% could be observed within the spectral range of interest. For the Gaussian noise, the standard error for the averaged measurement is proportional to $1/\sqrt{N}$, where $N$ is the number of samples. Thereby, the evaluated noise is verified, since the measurements coincide quite well with the results demonstrated in the previous section:
the obtained dependence yields the RMS of around 0.0010, derived for $N=3600$, while the measurements of the 100\% lines, where 4 spectra were averaged (900~pulses integrated for each spectrum), give a RMS of 0.00078.

\subsection{Quantitative measurements}

\begin{figure}[ht]
\begin{center}
\subfigure[Thermal source (130~um liquid flow cell)]{
\begin{tikzpicture}[remember picture]
  \node[anchor=south west,inner sep=0] (image) at (0,0,0) {\includegraphics[width=0.48\linewidth]{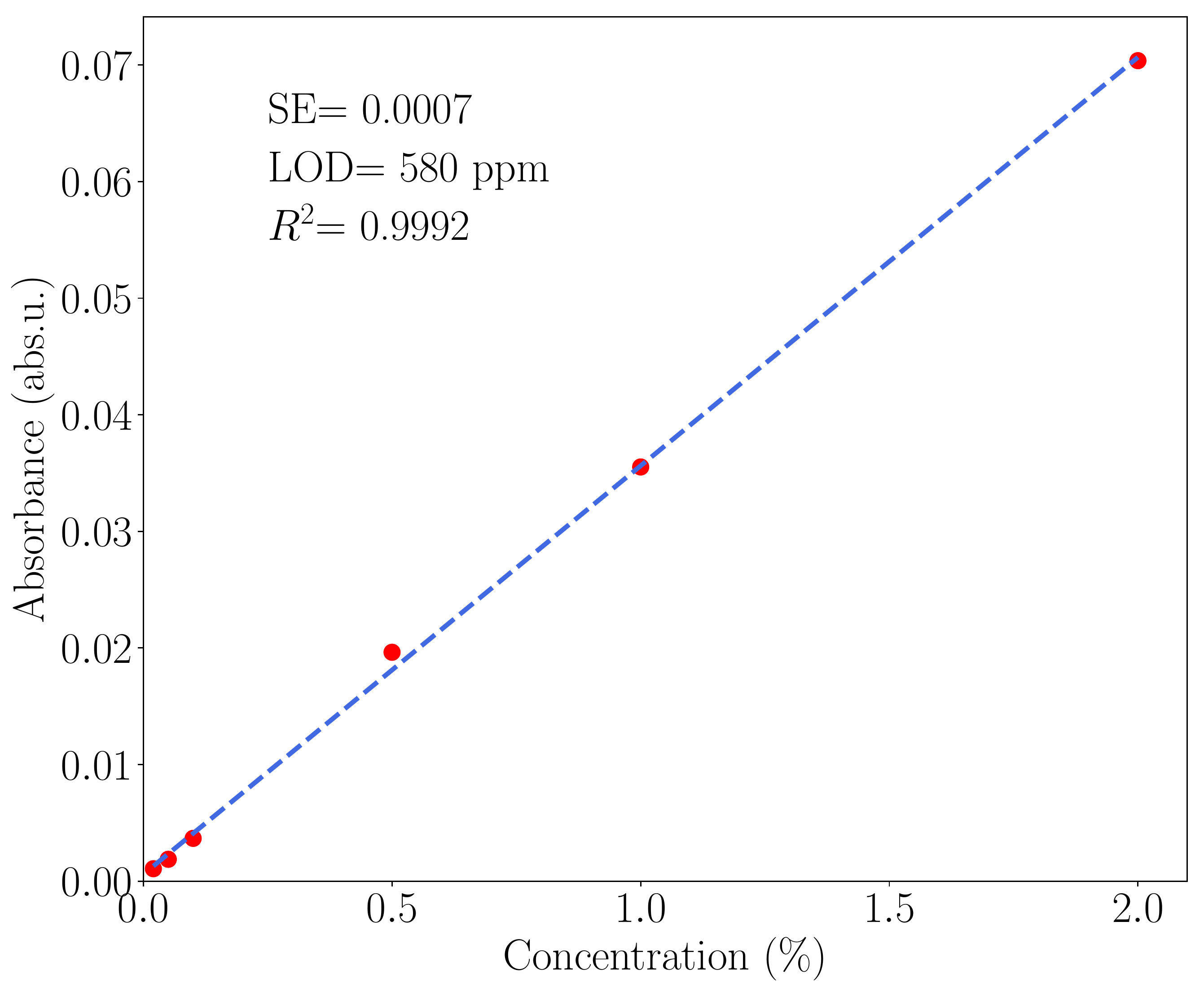}};
  \begin{scope}[x={(image.south east)},y={(image.north west)}]
  \end{scope}
\end{tikzpicture}
\label{fig:conctherm}}
\subfigure[Supercontinuum source (500~um liquid flow cell)]{
 \begin{tikzpicture}[]
  \node[anchor=south west,inner sep=0] (image) at (0,0,0) {\includegraphics[width=0.48\linewidth]{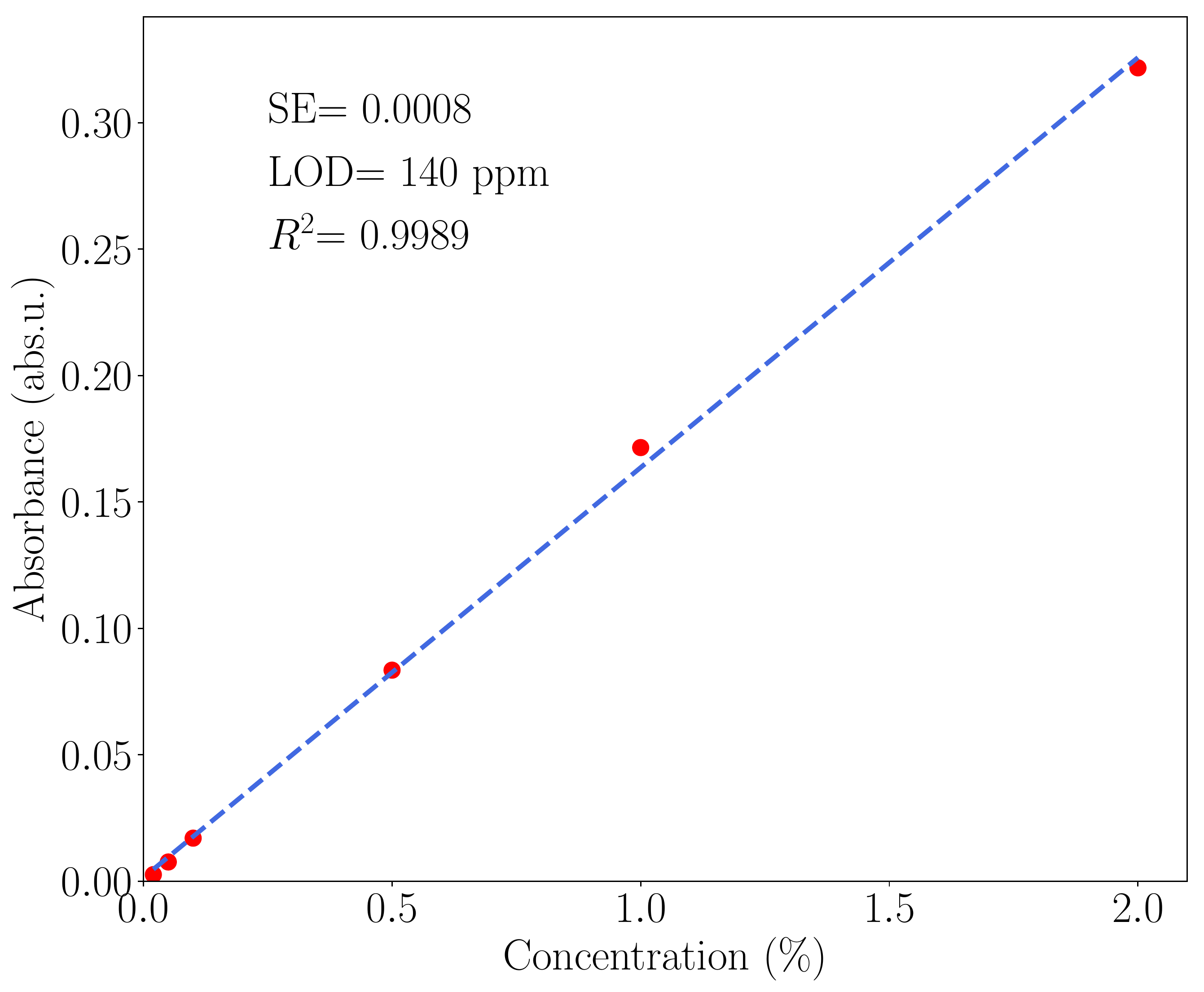}};
  \begin{scope}[x={(image.south east)},y={(image.north west)}]
  \end{scope}
\end{tikzpicture}
 \label{fig:concsc}}
    \begin{tikzpicture}[overlay,thick]
 			\node at (-1.85,2) {\includegraphics[scale=0.175]{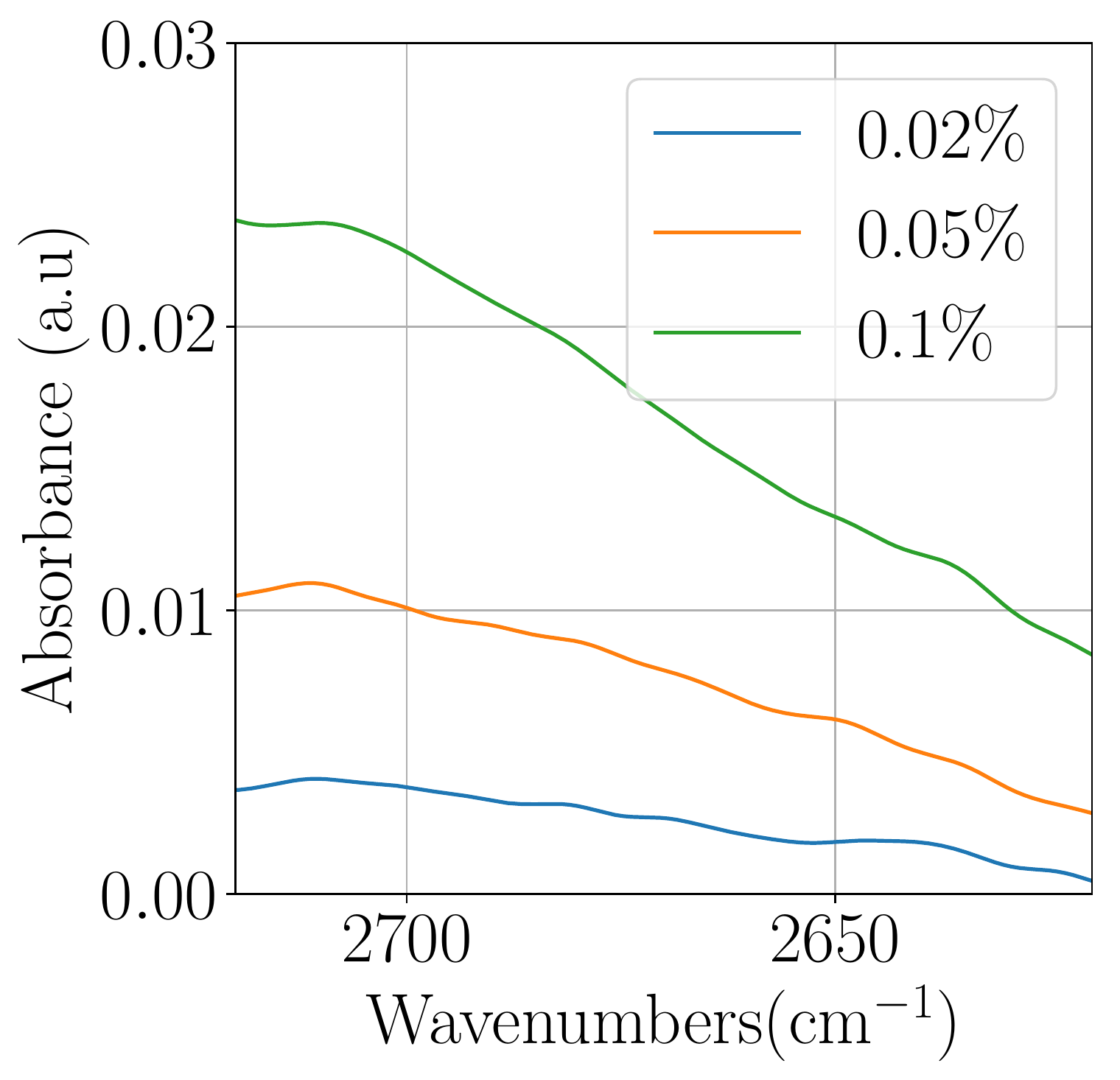}};

    \end{tikzpicture}
\caption{Calibration curves of a formaldehyde (methanediol) dilution series obtained by integrating the absorbance within the spectral range of 2720~cm\textsuperscript{-1} - 2620~cm\textsuperscript{-1}, covering parts of the C-H stretching band (a) standard FTIR instrument with CW thermal emitter and 130~\textmu m optical path length flow cell; (b) supercontinuum-based FTIR and 500~\textmu m flow cell.}
\label{fig:conse}
\end{center}
\end{figure}

The performance and analytical usability of the supercontinuum-based FTIR spectrometer were verified for the quantification of aqueous formaldehyde solutions (calibration curves shown in Fig.~\ref{fig:conse}).

The spectroscopic measurements were performed using both the CW thermal emitter (conventional system without external integrator) and the supercontinuum laser (experimental setup). The spacers of the liquid flow cell (130~\textmu m and 500~\textmu m) were selected correspondingly with respect to the available intensity levels, while the aperture size was set to 8~mm for both configurations. Despite the different path lengths, the raw signal recorded by FTIR in the case of the supercontinuum source exhibited a 50 times higher magnitude due to the distinctive properties of the radiation and the applied pulse integration approach; for each measurement 10 spectra were averaged.

The calculated absorbance spectra of the standard solutions were integrated within the spectral range of 2720~cm\textsuperscript{-1} - 2620~cm\textsuperscript{-1}, where the sufficient intensity is obtained for both light sources. Figure~\ref{fig:conse} depicts the obtained linear calibration curves. 
The inset in~Fig.~\ref{fig:concsc} presents absorbance spectra for the lowest concentrations of the dilution series. It should be noted that the spectral band, within which the spectroscopic analysis is performed, covers the range near the maximuma of the absorption band where it coincides with relatively weak water absorption.

The linear fitting model delivered $R^2$ values of 0.9992 for the thermal emitter and 0.9989 for the supercontinuum source respectively. Hence, corresponding slopes of 0.035 (thermal source) and 0.162 (supercontinuum) and standard errors (SE) yield the limits of detection (LOD) for both configurations calculated according to the IUPAC definition\ignore{\cite{lod1,lod2,BELTER2014606}}:\cite{lod2} a LOD of 580~ppm was achieved using conventional FTIR, while a superior LOD of 140~ppm has been demonstrated by the experimental system applying the supercontinuum source.

\section{Conclusions and Outlook}

Novel commercially available noise reduced mid-infrared (MIR) supercontinuum sources have recently evolved to a state where they can reasonably be applied for various spectroscopic analytical tasks\cite{KilgusApS:18,KilgusOE:18,Borondics:18,amiot,Gasser:18}. However, they are widely underestimated among spectroscopists, which is, for instance, illustrated by the fact that they are still basically unnoticed at the relevant conferences. The main reason might be a specification lag, i.e. high noise and instability, of their first versions. Therefore, one of the main goals of this contribution was to demonstrate that in fact MIR supercontinuum laser sources have achieved a level of maturity to be competitive with the state-of-the-art equipment. 
In order to support this idea, we demonstrated their applicability for the gold standard in MIR spectroscopy, Fourier-transform infrared spectroscopy (FTIR) that is strongly sensitive to intensity fluctuations. 
The basic idea here was to replace the broadband but weak and spatially incoherent thermal emitters by a broadband and high brightness spatially coherent source.

In the experimental part, a simple solution to overcome the sampling problem for the direct coupling of pulsed supercontinuum and FTIR has been proposed and realized. Utilizing the prospected experimental system, we achieved a factor of 200 greater amplitude of the raw interferogram and enhanced signal-to-noise ratio. The setup and the supercontinuum laser source were characterized with respect to noise and stability of the measurements. A satisfying long-term stability over around 28 hours and a noise level, which is comparable to the conventional thermal source, were demonstrated. The standard errors of the 100\% zero absorbance lines (0.67$\times$10\textsuperscript{-3} and 0.78$\times$10\textsuperscript{-3} for thermal and supercontinuum sources respectively) were calculated for the measurements of blank water within a 130~\textmu m liquid flow cell. 

The superior brightness of the supercontinuum source that is provided by the directionality of the radiation and the high output power allowed us to increase the interaction path length in a transmission measurement and to demonstrate the enhanced performance of the supercontinuum-based system. A spectroscopic analysis of an aqueous formaldehyde dilution series has been performed. The absorbance spectra (measured for 500~\textmu m liquid flow cell) yielded a limit-of-detection (LOD) of 140 ppm. In total, a 4-times enhanced detectivity over a conventional FTIR system with the thermal source was demonstrated, while the investigation of path lengths over 130~\textmu m were not expedient with the thermal emitter due to the almost total absorption by water.

The next generation of supercontinuum sources being developed is based on chalcogenide fibers and has already reached a milestone in terms of spectral coverage, starting in the NIR spectral region and spanning up to 14-16~\textmu m emission wavelength\cite{Dai_2018,Martinez:18,Ou:16,OB133,16um}. Thus, they cover almost the entire MIR spectral range, which makes them a highly promising technology to significantly push the field of MIR spectroscopy in the nearest future.

Considering the gained results and current developments\cite{Grassani:19,Jiao:19}, we believe that the proposed solution offers new potentials for enhancing the currently applied methods in this field. The demonstrated approach to couple pulsed supercontinuum sources appears to be universal and could be applied for any low-duty cycle source. Meanwhile, a purpose-oriented adaptation of the presented signal acquisition scheme or investigation of other types of integration devices, e.g. boxcar, could be considered in order to assemble a fully-integrated FTIR system, since a price reduction of supercontinuum sources is expected.

\section{Funding}

Financial support was provided by the Marie Sklodowska-Curie Action SUPUVIR of the European Union’s H2020-MSCA-ITN-2016 Programme under REA grant agreement no. 722380; and the strategic economic and research program ''Innovative Upper Austria 2020'' of the province of Upper Austria.

\bibliographystyle{unsrt}  

\end{document}